\documentclass[10pt]{IEEEtran}
\usepackage{graphicx}
\usepackage{amsmath}
\usepackage{epstopdf}
\usepackage{booktabs,chemarrow,multirow}
\usepackage{caption}
\usepackage{subfigure}
\usepackage[colorlinks,
 linkcolor=black,
 anchorcolor=black,
 citecolor=black]{hyperref}
\usepackage[all]{hypcap}

\begin{document}

\title{Performance Analysis of Reversible Binding \\Receptor Based Decode-and-Forward Relay \\in Molecular Communication Systems}

\author{Shuo Yuan,
        Jiaxing Wang,
        and~Mugen Peng,~\IEEEmembership{Senior~Member,~IEEE}
\thanks{Shuo Yuan (e-mail: yuanshchn@gmail.com), Jiaxing Wang (e-mail: jx19882008@163.com), and Mugen Peng (e-mail:
pmg@bupt.edu.cn) are with the Key Laboratory of Universal Wireless Communications (Ministry of Education), Beijing University of Posts and Telecommunications, Beijing 100876, China.}
\thanks{}
\thanks{}}

\maketitle

\begin{abstract}
Molecular communication (MC) allows nano-machines to communicate and cooperate with each other in a fluid environment. The diffusion-based MC is popular but is easily constrained by the transmit distance due to the severe attenuation of molecule concentrations. In this letter, we present a decode-and-forward (DF) relay strategy for the reversible binding receptor in the diffusion-based MC system. The time-varying spatial distribution of the information molecules based on the reversible association and dissociation between ligand and receptor at the surface of receiver is characterized. An analytical expression for the evaluation of expected error probability is derived, and the key factors impacting on the performance are exploited. Results show that with a constant molecular budget, the proposal can improve the performance significantly, and the performance gain can be enhanced by optimizing the position of the relay node and the number of molecules assigned to the source node.
\end{abstract}

\thispagestyle{empty}


\IEEEpeerreviewmaketitle

\section{Introduction}

\IEEEPARstart{D}{iffusion-based} molecular communication (DbMC) is considered as a particularly effective and energy-efficient approach of exchanging information among nano-machines \cite{Farsad2016Comprehensive}.  Unlike the active transport and bacterium-based communication, DbMC is a short-to-medium range molecular communication (MC) without external energy and infrastructure. Information molecules are encoded by transmitter propagation to the receiver based on free diffusion in DbMC\cite{Yilmaz2017Chemical}. During the propagation, however, the attenuation of molecular concentration worsens with the increasing distance. Thus, the reliable communication is challenging for the scenario of long transmit distance.

To solve this challenge, one potential solution inherited from the traditional wireless communication is to deploy relay between the transmitter and receiver. There have been several research efforts toward relay-assisted MC \cite{Qiu2017Bacterial}-[6]. The design and analysis of repeaters using bacterial has been investigated in \cite{Qiu2017Bacterial}, an information delivery energy model for MC via bacteria relays is established. In \cite{Wang2017performance} and \cite{Ahmadzadeh2015AmplifyandForward}, the authors propose a fixed-gain and variable-gain amplify-and-forward (AF) relay strategies. In \cite{Wang2015Relay}, a decode-and-forward (DF) relay strategy has been researched. However, most works assume that receiver node is transparent and the received signal is approximated by the local concentration of the information molecules inside the spherical receiver \cite{Ahmadzadeh2015AmplifyandForward, Wang2015Relay}. The reversible association and dissociation is a widely observed process for proteins and polymers. The performance characteristics of DF relay in a long-distance MC system using biological cells equipped with association and dissociation receptors should be researched to allow for the classification, optimization and realization of key techniques for internet of bio-nanothings \cite{Akyildiz2015internet}.

In this letter, a DF relay for diffusion-based MC is presented to improve performance in the long-distance scenario. The reversible binding receptor is able to associate a specific type of information molecule near its surface by receptor, and dissociates the molecules previously associated at its surface.
The novelties of this letter are summarized as follows.
\begin{itemize}
  \item A decode-and-forward molecular communication relay scheme for long-distance communication is concerned, in which an analytical model for the ligand-receptor reversible binding based DF relay in the diffusion-based MC system is formulated, and the Skellam distribution to approximate the number of ligand-receptor complex on the surface of receiver is presented.
  \item The corresponding error probability is characterized, and the key factors, such as the relay location, the rate of reversible binding and the number of molecules allocated to source node, impacting on the performance are exploited.
\end{itemize}

\section{System Model}

\begin{figure}[t] 
\centering
\includegraphics[width=0.45\textwidth]{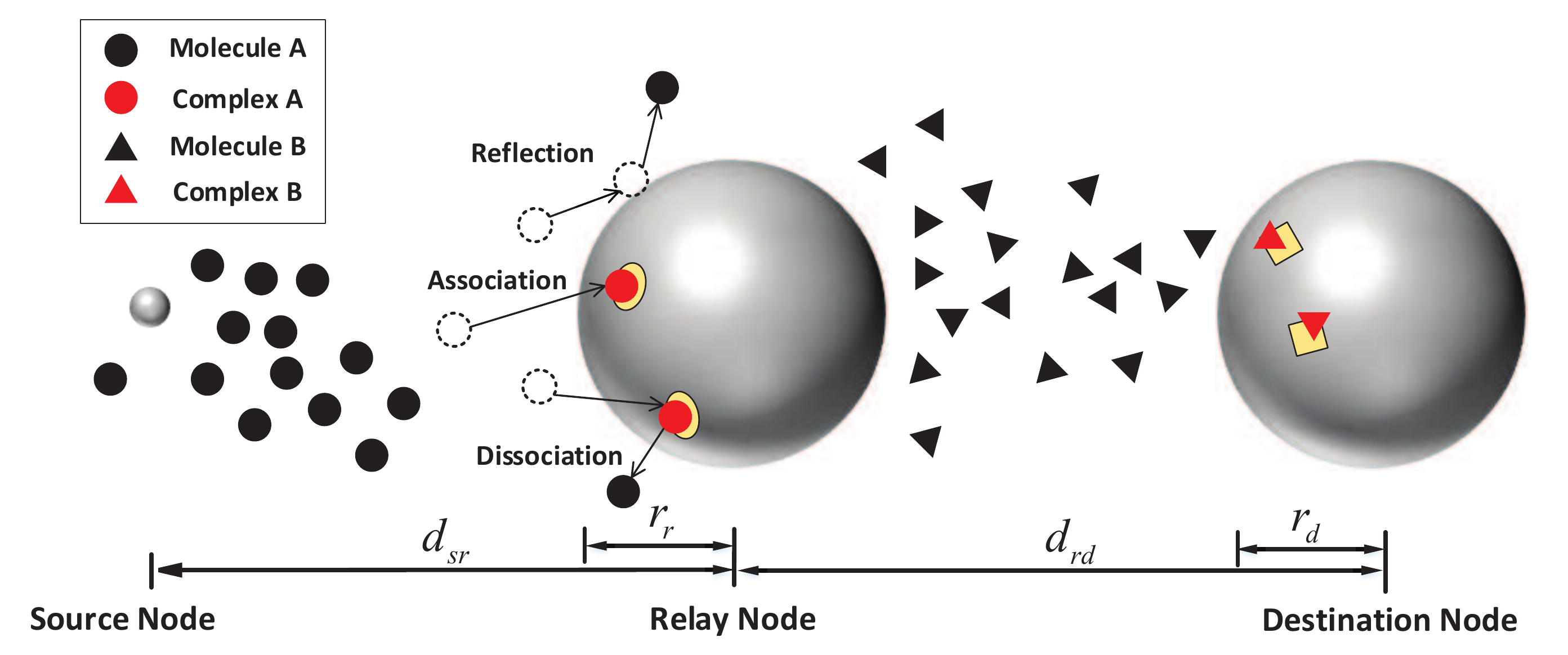}
\caption{Schematic diagram of considered system model. We assume that the spherical receiving (Relay node and Destination node) have no physical limitation on the number or placement of receptors.}
\label{system_model}
\end{figure}

We consider an unbounded three-dimensional (3D) DbMC system in a fluid environment without flow, consisting of a point transmitter (node S), a spherical receiver with radius $r_d$ (node D), and a spherical relay with radius $r_r$ (node R) used to extend the communication range, as illustrated in Fig. \ref{system_model}. The nodes S and D are placed at locations $\left( {0,0,0} \right)$ and $\left( {d_{sd},0,0} \right)$, respectively. Node R is spaced between node S and node D along the $x$-axis.

The on-off keying (OOK) is used, in which the transmission node (S or R) releases a fixed number of molecules at the beginning of a symbol duration to convey information bit ``1'' and releases nothing to transmit bit ``0''. The node R first decodes the received signal from node S, and then forwards it to node D. Furthermore, two different kinds of  non-interacting information molecules A and B are used at R for detection and re-transmission, respectively. Similarly with \cite{Deng2015Modeling}, the spherical receiving node (R or D) has no physical limitation on the number and the placement of receptors. Therefore, the number of molecules bound to the surface of receiving node (R or D) is not limited. This assumption is appropriate if the number of molecules is sufficiently low or the acceptor concentration is sufficiently high. It is assumed that the receptors on the surface of node R bind only to the type A molecules released by node S, and the receptors on the surface of node D bind only to the type B molecules released by node R. Once an information-carrying molecule binds to a receptor, a chain of chemical process is triggered to facilitate the counting of the molecules.

The reversible binding of information molecule (ligand) and receptor includes the transport step and reaction step. At the transport step, the transmission node (S or R) releases the processed molecules at the node center. Then, these molecules move independently over the fluidic medium according to Brownian motion to the receptors nearby. At the reaction step, the reflection, association and dissociation are independently executed. When the molecule contacts the receiver surface, it is either associated by the receptor to form ligand-receptor complex, or reflected into the fluid medium based on the association rate $k_{on}$ ($\mu {m}/s$). Based on the dissociation rate $k_{off}$ ($s^{-1}$), the ligand-receptor complex either dissociates or keeps it steady. When the dissociation occurs, the receptor releases the associated molecules to the fluid environment without changing its physical characteristics \cite{Deng2015Modeling}. The perfect synchronization among node S, node R and node D is assumed, and the reversible reaction can be described as:
\begin{equation}\label{reversible_reaction_description}
\begin{aligned}
L + E \autorightleftharpoons{$k_{on}$}{$k_{off}$} M.
\end{aligned}
\end{equation}
where $L$, $E$ and $M$ denote a ligand (information molecule), a receptor and a ligand-receptor complex, respectively.

To characterize the cumulative number of ligand-receptor complex at node R during the interval $[0,t]$, the cumulative reversible binding rate can be expressed as \cite{Deng2015Modeling}
\begin{equation}\label{cumulative_rate_reaction}
\begin{aligned}
\psi ({\Omega _{{r_r}}}&,t|{d_{sr}}) = 4{r_r}{D_A}\bigg(\int_0^\infty  {\frac{{\sin z}}{z}\operatorname{Re} } \left[ {U\left( {\frac{z}{t}} \right)} \right]dz + \\&\int_0^\infty  {\frac{{\cos z}}{z}\operatorname{Im} } \left[ {U\left( {\frac{z}{t}} \right)} \right]dz - \int_0^\infty  {\frac{1}{w}\operatorname{Im} } \left[ {U\left( w \right)} \right]dw\bigg),
\end{aligned}
\end{equation}
where ${D_A}$ is the diffusion coefficient, $\Omega _{{r_r}}$ denotes the surface area of node R, $d_{sr}$ is the distance between node S and the center of node R, and $U\left( {w} \right)$ can be expressed by:
\begin{equation}\label{U}
\begin{aligned}
U\left( {w} \right) &= \frac{1}{{4\pi {d_0}{D_A}}}\left( {1 - \frac{{\sqrt {\frac{{jw}}{{{D_A}}}} }}{{\left( {\frac{1}{{{r_r}}} + \frac{{{k_{on}}jw}}{{{D_A}\left( {jw + {k_{off}}} \right)}} + \sqrt {\frac{{jw}}{{{D_A}}}} } \right)}}} \right)
\\& \quad\times \exp \left\{ { - \left( {{d_{sr}} - {r_r}} \right)\sqrt {\frac{{jw}}{{{D_A}}}} } \right\}.
\end{aligned}
\end{equation}

Due to the independent Brownian movement of all information molecules, the time of hit the surface of receiver is random and can span multiple time slots. With the certain reversible binding rate which is derived in (\ref{cumulative_rate_reaction}), the number of molecules bound by receptor at the surface of node R during the interval $[0,n{T_b}]$ can be approximately modeled as a Poisson distribution:
\begin{equation}\label{cumulative_molecules_model_T}
\begin{aligned}
N_{s,r}^A\left( {0,n{T_b}} \right)\sim P\left( {{\lambda _{1s,r}}} \right),
\end{aligned}
\end{equation}
where $\lambda _{1s,r}$ is the mean value of $N_{s,r}^A\left( {0,n{T_b}} \right)$ as:
\begin{equation}\label{distribution_variance1}
\begin{aligned}
{\lambda _{1s,r}} = \sum\limits_{i = 1}^n {{N_A}{x_s}\left[ i \right]\psi \left( {{\Omega _{{r_r}}},\left( {n - i + 1} \right){T_b}|{d_{sr}}} \right)} ,
\end{aligned}
\end{equation}
and $N_A$ is the number of molecules released by node S to convey information bit ``1'', ${{x_s}\left[ i \right]}$ is the $i^{th}$ information bit transmitted by node S, and $T_b$ represents the bit interval.

Similarly, we can rewrite $N_{s,r}^A\left( {0,(n - 1){T_b}} \right)$ as following:
\begin{equation}\label{cumulative_molecules_model_T+Tb}
\begin{aligned}
N_{s,r}^A \left( {0,(n - 1){T_b}} \right) \sim P\left( \lambda _{2s,r}\right),
\end{aligned}
\end{equation}
where $\lambda _{2s,r}$ is the mean value of $N_{s,r}^A \left( {0,(n - 1){T_b}} \right)$ as:
\begin{align}
{\lambda _{2s,r}} &= \sum\limits_{i = 1}^{n - 1} {{N_A}{x_s}\left[ i \right]\psi \left( {{\Omega _{{r_r}}},\left( {n - i} \right){T_b}} |{d_{sr}}\right)}.
\label{distribution_variance2}
\end{align}

Based on (\ref{cumulative_molecules_model_T}) and (\ref{cumulative_molecules_model_T+Tb}), the number of molecules reversible binding by R at the $n^{th}$ time slot can be approximately modeled as the difference between two Poisson distributions:
\begin{equation}\label{receiver_distribution}
\begin{aligned}
N_{s,r}^A\left[ n \right] \sim P\left( {{\lambda _{1s,r}}} \right) - P\left( {{\lambda _{2s,r}}} \right),
\end{aligned}
\end{equation}

Note that the dependence between $N_{s,r}^A\left( {0,n{T_b}} \right)$ and $N_{s,r}^A\left( {0,(n - 1){T_b}} \right)$ can be ignored for a sufficiently big bit interval. The difference $N_{s,r}^A\left[ n \right]$ between two Poisson distribution $N_{s,r}^A\left( {0,n{T_b}} \right)$ and $N_{s,r}^A\left( {0,(n - 1){T_b}} \right)$ with mean values $\lambda _{1s,r}$ and $\lambda _{2s,r}$ can be described as a Skellam distribution. Thus, (\ref{receiver_distribution}) can be rewritten as:
\begin{equation}\label{skellam_distribution}
\begin{aligned}
N_{s,r}^A\left[ n \right] \sim skellam\left( {{\lambda _{1s,r}},{\lambda _{2s,r}}} \right),
\end{aligned}
\end{equation}

The probability mass function for the Skellam distribution $N_{s,r}^A\left[ n \right]$ is given by:
\begin{equation}\label{skellam_distribution_pmf}
\begin{aligned}
\rho &(m;{\lambda _{1s,r}},{\lambda _{2s,r}}) = \Pr \left\{ {N_{s,r}^A\left[ n \right] = m} \right\} \\&= \exp \left\{ { - \left( {{\lambda _{1s,r}} + {\lambda _{2s,r}}} \right)} \right\}{\left( {\frac{{{\lambda _{1s,r}}}}{{{\lambda _{2s,r}}}}} \right)^{\frac{m}{2}}}{I_m}\left( {2\sqrt {{\lambda _{1s,r}}{\lambda _{2s,r}}} } \right).
\end{aligned}
\end{equation}
where ${I_m}\left( \cdot \right)$ denotes the modified Bessel function of the first kind.

According to (\ref{cumulative_molecules_model_T})-(\ref{skellam_distribution_pmf}), the number of B molecules bound by receptor of node D at the ${\left( {n + 1} \right)^{th}}$ time slot can be expressed as same as the calculation of the number of A molecules bound by receptor of node R, which is denoted by $N_{r,d}^B\left[ n+1 \right]$. Thus, the distribution of $N_{r,d}^B\left[ n+1 \right]$ can be written as:
\begin{equation}\label{r_d_skellam_distribution}
\begin{aligned}
N_{r,d}^B\left[ n+1 \right] \sim skellam\left( {{\lambda _{1r,d}},{\lambda _{2r,d}}} \right),
\end{aligned}
\end{equation}
where the mean values are given by:
\begin{equation}\label{la1rd}
\setlength{\abovedisplayskip}{2pt}
\setlength{\belowdisplayskip}{2pt}
\begin{aligned}
{\lambda _{1r,d}} = \sum\limits_{i = 1}^{n + 1} {N_B}{x_r}\left[ i \right]\psi \left( {{\Omega _{{r_d}}},\left( {n - i + 2} \right){T_b}} |{d_{rd}}\right) ,
\end{aligned}
\end{equation}
\begin{equation}\label{la2rd}
\begin{aligned}
{\lambda _{2r,d}} = \sum\limits_{i = 1}^n {N_B}{x_r}\left[ i \right]\psi \left( {{\Omega _{{r_d}}},\left( {n - i + 1} \right){T_b}} |{d_{rd}}\right),
\end{aligned}
\end{equation}
where $N_B$ is number of molecules released by node R to convey information bit ``1'', ${{x_r}\left[ i \right]}$ is the $i^{th}$ information bit transmitted by node R, $\Omega _{{r_d}}$ denotes the surface area of node D, and $d_{rd}$ represents the distance between the centers of node R and node D.

\section{Performance Analysis}

For the signal detection, the receiver can apply maximum-a-posterior (MAP) to decide the received information, which can be expressed as:
\begin{equation}\label{MAP}
\begin{aligned}
{\hat y_r}\left[ n \right] = \left\{ {\begin{array}{*{20}{c}}
{1\;\;\;\;{\rm{if }}\;\;\;{\rm{ }}N_{s,r}^A\left[ n \right] \ge {\tau _R}}\\
{0\;\;\;\;{\rm{if }}\;\;\;N_{s,r}^A\left[ n \right] < {\tau _R}}
\end{array}} \right.,
\end{aligned}
\end{equation}
where ${\tau _R}$ is the detection threshold at node R, and ${\hat y_r}\left[ n \right]$ is the information bit detected by node R in the $n^{th}$ time slot. Node R is assumed to re-encode ${\hat y_r}\left[ n \right]$ accurately, and forwards to node D at the beginning of the ${\left( {n + 1} \right)^{th}}$ time slots after the successful decoding, which is denoted by ${x_r}\left[ {n + 1} \right]$.

According to Skellam distribution and MAP detection method in (\ref{MAP}), the bit error probability of transmitting bit ``1'' from node S to node R at the ${n^{th}}$ time slot can be written as:
\begin{small}
\begin{equation}\label{BER_S1R0}
\begin{aligned}
\Pr[{\hat y_r}\left[ n \right] = 0|{x_s}\left[ n \right] = 1] &= \Pr (\left. {N_{s,r}^A\left[ n \right] < {\tau _R}} \right|{x_s}[n] = 1) \\&\approx \sum\limits_{m =  - \infty }^{{\tau _R} - 1} {\rho_1 (m;{\lambda _{1s,r}},{\lambda _{2s,r}})} ,
\end{aligned}
\end{equation}
\end{small}
where ${\rho_1 (m;{\lambda _{1s,r}},{\lambda _{2s,r}})}$ denotes the distribution of $N_{s,r}^A\left[ n \right]$ when node S sending the information bit ``1'' at the ${n^{th}}$ time slot, ${{\lambda _{1s,r}}}$ and ${{\lambda _{2s,r}}}$ are given in (\ref{distribution_variance1}) and (\ref{distribution_variance2}), respectively.

Analogously, the bit error probability of transmitting bit ``0'' from node S to node R at the ${n^{th}}$ time slot can be expressed as:
\begin{small}
\begin{equation}\label{BER_S0R1}
\begin{aligned}
\Pr\left[ {{{\hat y}_r}\left[ n \right] = 1|{x_s}\left[ n \right] = 0} \right] &= \Pr \left( {\left. {N_{s,r}^A\left[ n \right] \ge {\tau _R}} \right|{x_s}[n] = 0} \right)\\& \approx \sum\limits_{{\tau _R}}^{m = \infty } {\rho_0 (m;{\lambda _{1s,r}},{\lambda _{2s,r}})},
\end{aligned}
\end{equation}
\end{small}
where ${\rho_0 (m;{\lambda _{1s,r}},{\lambda _{2s,r}})}$ denotes the distribution of $N_{s,r}^A\left[ n \right]$ when node S sending the information bit ``0'' at the ${n^{th}}$ time slot.

The probability of sending information bit ``0'' and bit ``1'' are $\Pr \left( {{x_s}\left[ n \right] = 0} \right)=P_0$ and $\Pr \left( {{x_s}\left[ n \right] = 1} \right)=P_1$, respectively. In particular, $\Pr \left( {{x_r}\left[ n \right] = 0} \right)$ can be written as:
\begin{equation}\label{BER_r1}
\begin{aligned}
\Pr \left( {{x_r}\left[ n \right] = 0} \right)& = P_0 \times \Pr \left( {{{\hat y}_r}\left[ n \right] = 0\left| {{x_s}\left[ n \right] = 0} \right.} \right)\\&+ P_1 \times \Pr \left( {{{\hat y}_r}\left[ n \right] = 0\left| {{x_s}\left[ n \right] = 1} \right.} \right) .
\end{aligned}
\end{equation}

Similarly with the calculation of the bit error probability of the transmitted information bit from node S to node R as (\ref{BER_S1R0}) and (\ref{BER_S0R1}), the bit error probability of the transmitted information bit from node R to node D can be calculated.

The error probability of these two-hops for the ${n^{th}}$ bit can be written as:
\begin{equation}\label{pe_sim}
\begin{aligned}
{\rm{Pe}}[n] &= P_1 \times \Pr \left( {{{\hat y}_d}\left[ {n + 1} \right] = 0\left| {{x_s}\left[ n \right] = 1} \right.} \right) \\&+ P_0 \times \Pr \left( {{{\hat y}_d}\left[ {n + 1} \right] = 1\left| {{x_s}\left[ n \right] = 0} \right.} \right),
\end{aligned}
\end{equation}
where ${{\hat y}_d}\left[ {n + 1} \right]$ is the information bit detected by node D in the $(n+1)^{th}$ time slot.

Let ${p_{s0r0}}[n] = \Pr \left( {{{\hat y}_r}\left[ n \right] = 0\left| {{x_s}\left[ n \right] = 0} \right.} \right)$. Similarly, ${p_{s0r1}}[n]$, ${p_{s1r0}}[n]$, and ${p_{s1r1}}[n]$ can be derived. The probability of the cases in the second hop can be expressed as ${p_{r0d0}}[n+1]$, ${p_{r0d1}}[n+1]$, ${p_{r1d1}}[n+1]$ and ${p_{r1d0}}[n+1]$, respectively. According to the chain rule, the first term on the right side of (\ref{pe_sim}) can be derived as:
\begin{equation}\label{ber_s1d0}
\begin{aligned}
\Pr &\left( {{{\hat y}_d}\left[ {n + 1} \right] = 0\left| {{x_s}\left[ n \right] = 1} \right.} \right)\\
 &= {p_{s1r1}}[n] \times {p_{r1d0}}[n+1] + {p_{s1r0}}[n]\times{p_{r0d0}}[n+1],
\end{aligned}
\end{equation}

As a consequence, formula (\ref{pe_sim}) can be derived as:
\begin{equation}\label{ber_two_hop}
\begin{aligned}
{\text{Pe}}&[n]{\text{ }} \\&= {P_1}({p_{s1r1}}[n] \times {p_{r1d0}}[n + 1] + {p_{s1r0}}[n] \times {p_{r0d0}}[n + 1]) \\&+ {P_0}({p_{s0r0}}[n] \times {p_{r0d1}}[n + 1] + {p_{s0r1}}[n] \times {p_{r1d1}}[n + 1]).
\end{aligned}
\end{equation}

Considering the special case of $P_0 = P_1 = \frac{1}{2}$, the error probability of two-hop for $n^{th}$ bit can be derived as:
\begin{equation}\label{ber_pe_sim}
\begin{aligned}
&{\rm{2Pe}}[n] \\&= \bigg[\sum\limits_{m =  - \infty }^{{\tau _R} - 1} {{\rho _0}(m;{\lambda _{1s,r}},{\lambda _{2s,r}})}  \times \sum\limits_{{\tau _D}}^{m = \infty } {{\rho _0}(m;{\lambda _{1r,d}},{\lambda _{2r,d}})} \\& + \sum\limits_{{\tau _R}}^{m = \infty } {{\rho _0}(m;{\lambda _{1s,r}},{\lambda _{2s,r}}) \times \sum\limits_{{\tau _D}}^{m = \infty } {{\rho _1}(m;{\lambda _{1r,d}},{\lambda _{2r,d}})} } \bigg]\\& + \bigg[\sum\limits_{{\tau _R}}^{m = \infty } {{\rho _1}(m;{\lambda _{1s,r}},{\lambda _{2s,r}}) \times \sum\limits_{m =  - \infty }^{{\tau _D} - 1} {{\rho _1}(m;{\lambda _{1r,d}},{\lambda _{2r,d}})} } \\& + \sum\limits_{m =  - \infty }^{{\tau _R} - 1} {{\rho _1}(m;{\lambda _{1s,r}},{\lambda _{2s,r}})}  \times \sum\limits_{m =  - \infty }^{{\tau _D} - 1} {{\rho _0}(m;{\lambda _{1r,d}},{\lambda _{2r,d}})} \bigg].
\end{aligned}
\end{equation}

\section{Numerical Results}

\begin{figure}[t] 
\centering
\includegraphics[width=0.45\textwidth]{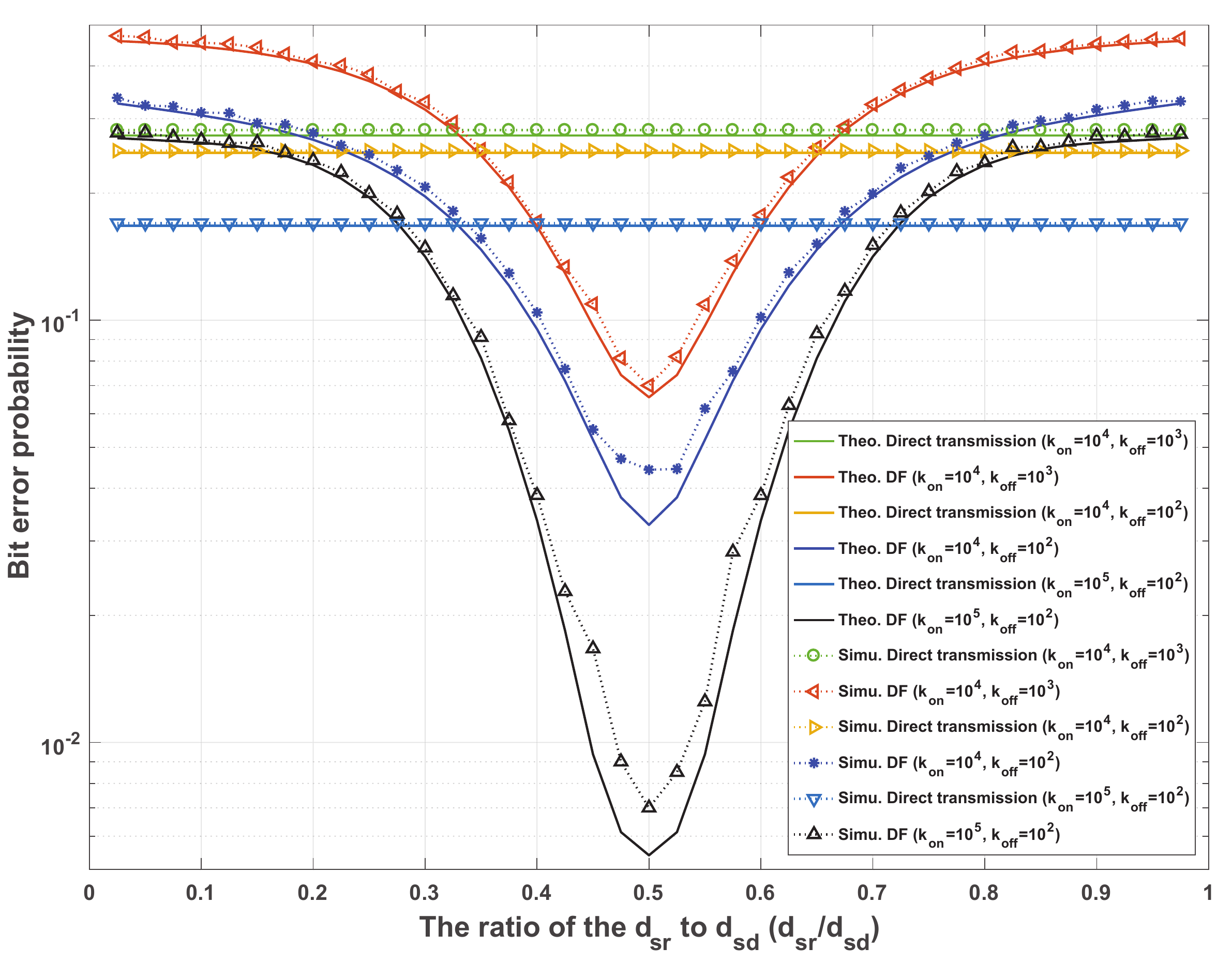}
\caption{Bit error probability as a function of the ratio of the $d_{sr}$ to $d_{sd}$ ($N_A = N_B = 1000$, $N_{A(direct)} = 2000$, $d_{sd} = 30 {\mu}m$)}
\label{fig1_1}
\end{figure}

In this section, the stochastic simulation framework proposed in \cite{Deng2015Modeling} is extended. For all cases, the simulation results are averaged over $10^4$ independent snapshots. A third bit sequence used in the simulation, where the first 2 bits are ``1 1'' and the last bit is bit ``1'' and bit ``0'', respectively. The units for the association rate $k_{on}$ and dissociation rate $k_{off}$ are ${\mu}m/s$ and $s^{-1}$, respectively. Some key factors are set as following: $r_r = r_d = 5 {\mu}m$, $D_A = 79.4 {\mu}m^2/s$, $T_b = 0.7 s$, the distance between node S and node D $d_{sd} = 30 {\mu}m$, and the sampling interval $T_s = 0.002 s$ \cite{Deng2015Modeling}. Furthermore, the molecules of type A and type B have the same diffusion coefficient in the fluid environment. The molecular budget of the system keeps constant $N_A + N_B = 2000$.

Fig. 2 presents the minimum error probability that can be achieved for each position of node R. At each position of node R, the bit error probability versus different decision threshold for both nodes R and D is obtained, and then the optimal decision threshold with the minimal bit error probability is approached. Firstly, it can be observed that the theoretical curves match well the simulation curves, which shows the accurateness of the theoretical analysis. In addition, the concerned DF relay can improve the performance significantly with respect to the direction transmission in our model. For the same $k_{on}$, the quality of the communication improves with decreasing $k_{off}$. This happens because the received signal for bit ``1'' is more distinguishable than that for bit ``0'' by decreasing $k_{off}$. On the other hand, for the same $k_{off}$, the quality of the communication degrades with decreasing $k_{on}$, which is because the received signal for bit ``1'' is less distinguishable than that for bit ``0'' with decreasing $k_{on}$. Furthermore, it is clearly illustrated that with the increasing distance between the relay location and the middle point, the errors grow up. Especially, when the ratio $(d_{sr}/d_{sd})$ is sufficiently large or small, the relay may not be helpful for performance improvement. This happens because when $d_{sr}$ is far away from $d_{rd}$, the number of molecules released by nodes S and D is not properly assigned.

\begin{figure}[t] 
\centering
\includegraphics[width=0.45\textwidth]{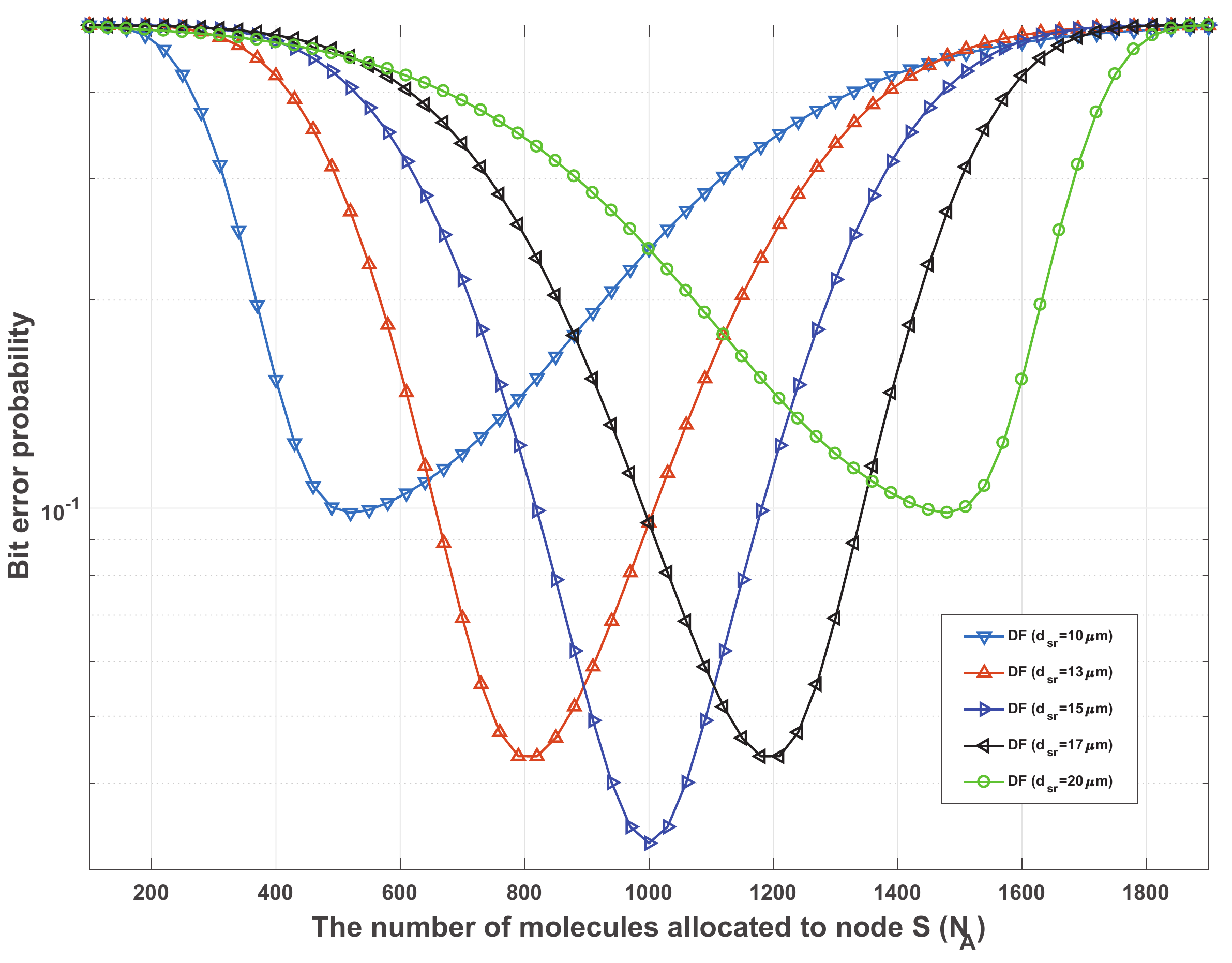}
\caption{Bit error probability of the DF relay-assisted MC system versus the number of molecules allocated to node S for different positions of node R ($N_A + N_B = 2000$, $k_{on} = 10^4$, $k_{off} = 100$, $d_{sd} = 30 {\mu}m$)}
\label{fig1_1}
\end{figure}

For different positions of the node R, the quality of the communication with respect to the number of molecules assigned to node S is shown in Fig. 3. At each position of node R, as the number of molecules assigned to node S increases, the bit error probability significantly decreases until the minimum point is reached, and then the bit error probability increases with the increasing number of molecules. As a result, there is an optimal molecular number, which can yield the minimum error probability for each position of node R. Moreover, the best performance of the proposed DF relay for MC can be achieved in the case of $N_A = N_B$ and $d_{sr} = d_{rd}$.

\section{Conclusion}

In this letter, an analytical model for the ligand-receptor reversible binding based decode-and-forward (DF) relay molecular communication (MC) has been proposed. The number of ligand-receptor complex has been approximately modeled as a Skellam distribution, and the corresponding bit error probability has been theoretically derived and evaluated. The key factors impacting on the MC performance have been exploited as well. The analysis and simulation results have shown that the performance of ligand-receptor reversible binding based DF relay MC can be significantly improved by increasing the ratio of association to dissociation, optimizing the number of molecules distribution scheme, or assigning an optimal location of relay node. In the future, the machine learning can be used for the channel \cite{Lee2017Machinea}, and some new performance metrics can be researched in the DF relay MC systems.





\ifCLASSOPTIONcaptionsoff
 \newpage
\fi

\bibliographystyle{IEEEtran}
\bibliography{Peng_WCL2018-0208}

\end{document}